\documentclass{aa}  

\usepackage{graphicx}
\usepackage{txfonts}
\defcitealias{PerriLeitner2022}{Perri, Leitner, et al., 2022} 

\begin{document}

   \title{COCONUT-MF: Two-fluid ion-neutral global coronal modelling}


   \author{M. Brchnelova
          \inst{1}
          \and
          B. Ku\'zma\inst{2} \and
          F. Zhang \inst{1} \and
          A. Lani \inst{1} \and 
          S. Poedts \inst{1,3}
          }

   \institute{Centre for Mathematical Plasma Astrophysics, KU Leuven,
              Celestijnenlaan 200B, 3001, Leuven\\
              \email{michaela.brchnelova@kuleuven.be}
         \and
             Shenzhen Key Laboratory of Numerical Prediction for Space Storm, Institute of Space Science and Applied Technology, Harbin Institute of Technology, Shenzhen, People's Republic of China, 518055 
        \and Institute of Physics, University of Mariia Curie-Sk{\l}odowska, ul.\ Radziszewskiego 10, 20-031 Lublin, Poland 
             }

   \date{Received ? ??, ????; accepted 2 August, 2023}

 
  \abstract
   {The global coronal model COCONUT (COolfluid COronal uNstrUcTured) was originally developed to replace semi-empirical models such as the Wang-Sheeley-Arge model in space weather forecasting chains in order to improve the physical accuracy of the predictions. This model has, however, several simplifications implemented in its formulation to allow for rapid convergence in an operational setting. These simplifications include the assumptions that the plasma is fully ionised, sufficiently collisional, and that quasi-neutrality holds, so that it can be modelled as a single fluid. This means that all interactions with the low-concentration neutral fluid in the corona, such as collisions or charge exchange, are neglected.}
   {In this paper, we have two goals. Firstly, we aim to introduce a novel multi-fluid global coronal model and validate it with simple cases (like a magnetic dipole) as well as with real data-driven applications. Secondly, we aim to investigate to what extent considering a single-fluid plasma in the global coronal model might affect the resulting plasma dynamics, and thus whether the assumptions on which the single-fluid coronal model is based are justified.}
   {We developed a multi-fluid global coronal model following the ideal magnetohydrodynamics (MHD) COCONUT model, COCONUT-MF, which resolves the ion and neutral fluid equations separately. While this model is still steady-state and thus does not resolve unsteady processes, it can account for resistivity, charge exchange, and chemical (ionisation and recombination) and collisional contributions due to the presence of the neutrals in the fluid equations.}
   {We present the results of the ion-neutral COCONUT-MF modelling for a magnetic dipole, a minimum of solar activity case (August 1, 2008), and a solar maximum case (March 9, 2016). Through comparison with the ideal MHD results, we confirm that the resolved multi-fluid solver features are physical and also demonstrate the higher accuracy of the applied upwind numerical flux scheme compared to the one used in the original MHD model. Subsequently, we also repeat the multi-fluid simulations while excluding the charge exchange and the chemical and collisional terms to evaluate the effect these terms have on the resulting plasma dynamics. It is observed in numerical results that, despite the very low concentration of neutrals, they still do affect the flow field to a limited but non-negligible extent (up to 5 to 10\% locally), with a higher impact being seen in the case of the solar maximum. It is also demonstrated that the collisional terms are primarily responsible for the neutrals adopting the electromagnetic profiles of the ions, while the charge exchange and chemical terms yield the largest thermal effects of the neutrals on the ion plasma. Despite the fact that the coronal plasma is generally assumed to be collisionless, our results show that there is sufficient collisionality in it to couple the two fluids. 
   }
   {We present a novel multi-fluid global coronal model that can separately simulate the behaviour of the ion and neutral fluids. Using this model, we also show that in our set-up, in which the chromosphere is not considered and steady-state solutions are assumed, the presence of the neutrals affects the flow field, though to a limited extent. It is shown that this effect is larger when the flow field is more complex due to a higher magnetic activity. This analysis may change in the future when the global coronal model will be extended to include the lower atmospheric layers as well as terms to model coronal heating, radiation, and thermal conduction. To that end, the current model may need to be further calibrated to better represent the different layers of the atmosphere. We presume that the use of the proposed COCONUT-MF set-up will then be necessary and new numerical experiments will need to be performed in order to confirm this hypothesis. }


   \maketitle
%

\section{Introduction}

The forecasting of multiple space weather effects has been receiving growing interest in the past few decades as our society is relying increasingly on modern electronics, digital technologies, and space-borne infrastructure, which are susceptible to possibly severe geomagnetic storms. In order to improve space weather forecasts, a new global coronal model, COolfluid COronal uNstrUcTured, aka COCONUT \citepalias{PerriLeitner2022}, has recently been developed. This code is based on ideal magnetohydrodynamics (MHD), utilising the COOLFluiD framework \citep{Lani2005, Kimpe2005, Lani2013, Lani2014} due to its implicit formulation and the use of unstructured grids. Over the course of testing, COCONUT was shown to produce results with remarkable speed and good accuracy, even in the more challenging cases during solar maximum, recreating most of the observational electromagnetic features successfully when comparing the resolved magnetic field to white-light images from solar eclipses \citep{Kuzma2023}. These results can be obtained in just a few hours when running on a supercomputer, making COCONUT simulations operationally feasible for space weather forecasts. {This code can then be run in combination with other assets such as EUHFORIA \citep{Pomoell18} or ICARUS \citep{Verbeke2022}, where COCONUT is able to compute the domain typically from 1$\;R_\odot$ to 0.1 AU and the heliospheric codes EUHFORIA and ICARUS use the data from COCONUT at their inner boundaries of 0.1 AU to compute the inner heliospheric conditions from 0.1 AU onwards. These couplings can then be utilised within space weather forecasting tool chains, such as those implemented in the ESA Virtual Space Weather Modelling Centre \citep{vswmc}. }


As of now, the current version of COCONUT only includes a relatively simple model with reduced physics complexity, assuming a polytropic plasma without appropriate terms in the equations to describe coronal heating, radiative losses, and thermal conduction. Work is currently ongoing to include these terms in the formulation using a variety of approximations so that a solar wind solution with a more realistic velocity profile can be obtained. Another simplifying assumption made in COCONUT is that the plasma is described within the framework of a single-fluid MHD model, such as in the majority of state-of-the-art global coronal models. While at temperatures of 1M to 2M Kelvins, it is indeed expected that the plasma is very close to being fully ionised, some small amount of neutrals are still present, and ionisation, recombination, and charge exchange are occurring continuously. Including and studying these effects would require at least a two-fluid model. The effects of two-fluid phenomena in partially ionised plasma of the solar atmosphere have already been investigated by several authors \citep[e.g.][]{Zaqarashvili2011,Tu_2013,Khomenko2014,Popescu2019,Wojcik2019,Murawski2020,Kuzma2021a,Kuzma2021b,Zhang2021}. These studies have shown that two-fluid effects cannot be neglected in the solar chromosphere, where the plasma is denser and the ionisation degree is lower. It was further shown that, even within the hot corona, regions of locally concentrated, colder plasma are prone to two-fluid effects, for example in prominences \citep{Wiehr2021} and coronal rain blobs \citep{MartinezGomez2022}. Whether the very small concentration of neutrals can have an impact on the global dynamics of the ions in the corona has, to our knowledge, never been investigated and, so far, it has been taken for granted that they do not have any effect. However, even though the concentration of neutrals is very small, especially the momentum and energy source terms originating from the two-fluid effects under coronal conditions could be large enough to affect the ion temperature and velocity, at least locally. It is thus one of the main purposes of this paper to take these effects into account and assess their impact.


The development of a multi-fluid (MF) ion-neutral global coronal formulation is, however, important for us even beyond this rationale. It is our long-term aim to eventually develop a version of COCONUT that would extend down to the lower solar atmosphere and include the transition region and the chromosphere, such that we can better capture the corresponding physics, including the energy transfer between the different layers of the solar atmosphere, and approximate the coronal heating in more detail and, hence, improve the overall accuracy of the results. Especially in the chromosphere, as shown in the short literature review above, due to the lower temperatures and higher density, the concentration and the effect of neutrals are definitely not negligible and thus a MF formulation will have to be applied. For that purpose, introducing the neutrals as a separate fluid into our set-up (as of now only containing the coronal domain, without the chromosphere) and validating the correct operation of the solver is an essential first step. Thus, the development and validation of this ion-neutral version of COCONUT is the second main aim of this paper. 

The remainder of this paper is organised as follows. First, the COCONUT-MF formulation and simulation set-up are described in Section~\ref{sec:setup}. Next, this formulation is applied to the simulation of a magnetic dipole in Section~\ref{sec:dipole}, in which a comparison with the ideal MHD solution is carried out and the differences are identified and explained. Next, the multi-fluid set-up is used to simulate data-driven cases. We consider both a solar minimum and a solar maximum case in Section~\ref{sec:applicationreal} in order to show the application to  data-driven and more physically realistic cases. A comparison with MHD is again performed, followed by an analysis of the significance of the different ion-neutral terms for the resulting behaviour of the plasma. The performance of the new solver from the perspective of convergence and computational resources required is elaborated on in Section~\ref{sec:performance}. Finally, Section~\ref{sec:conclusion} summarises the findings of the paper. 

\section{Multi-fluid MHD set-up}
\label{sec:setup}
%
Let us first introduce the ion-neutral formulation and the set-up of the newly developed COCONUT-MF solver.

\subsection{Formulation}

COOLFluiD as a framework has been used for chromospheric two-fluid modelling in the past (see e.g. \cite{AlvarezLaguna_2017}). For more details about the numerical aspects of the COOLFluiD multi-fluid solver that has been used in this study we refer to \cite{ALVAREZLAGUNA2016252}, which also shows a verification of the code on a number of test cases for a variety of plasma conditions.

In the formulae below, the following definitions are used:
   \[
      \begin{array}{lp{0.8\linewidth}}
        \mathbf{B} & magnetic field, \\
        c & speed of light, \\
         \mathbf{E} & electric field, \\
          \mathbf{g} & gravitational acceleration, \\
         \mathbf{J} & electric current,  \\
         m_i, m_n & mass of ions and neutrals, \\
         n_i, n_n & number density of ions and neutrals, \\
         \mathbb{P}_{i}, \mathbb{P}_{n} & ion and neutral pressure tensors, \\
         P & thermal pressure scalar, \\
         q & electric charge, \\
         T_i, T_n, T_e & temperature of ions, neutrals, and electrons ($T^*$ is given in eV), \\
         t & time, \\
         \mathbf{V}_i, \mathbf{V}_n & velocity of ions and neutrals, \\
         v & speed, \\
         \gamma & ratio of specific heats, \\
         \epsilon_0 & vacuum permitivity, \\
        \varepsilon_i, \varepsilon_n & total ion and neutral fluid energy, \\
         \mu_0 & vacuum magnetic permeability, \\
         \rho_i, \rho_n & mass density of ions and neutrals, \\
         \Sigma & collisional cross-section, \\
         \sigma_{t} & total charge, and \\
         \Psi, \Phi & Lagrange multipliers, \\
      \end{array}
   \]
   
\noindent  while other, less frequently used terms are explained in-text. The approximations below largely follow the derivations of \citet{Leake2012}, \citet{Leake2013}, \citet{MeierShumlak2012}, and \citet{Murawski_2022}.

The MF ion-neutral formulation (which we will sometimes refer to as MFMHD in the remainder of this paper) was implemented as follows. The Maxwell equations, with the hyperbolic divergence cleaning (HDC) method implemented in the solver \citep{MUNZ200083}, read

\begin{equation}
\frac{\partial \mathbf{B}}{\partial t}+{\nabla} \times \mathbf{E}+\gamma^2 \vec{\nabla} \Psi=0,
\label{eq:no1}
\end{equation}

\begin{equation}
\frac{\partial \mathbf{E}}{\partial t}-c^2 {\nabla} \times \mathbf{B}+(\chi c)^2 {\nabla} \Phi=-\frac{\mathbf{J}}{\mu_0},
\end{equation}

\begin{equation}
\frac{\partial \Psi}{\partial t}+c^2 {\nabla} \cdot \mathbf{B}=0,
\quad \text{and}
\end{equation}

\begin{equation}
\frac{\partial \Phi}{\partial t}+{\nabla} \cdot \mathbf{E}=\frac{\sigma_t}{\epsilon_0},
\end{equation}

\noindent where the two Lagrange multipliers of the HDC method are $\Psi$ and $\Phi$. The HDC method introduces artificial waves of speeds $\chi c$ and $\gamma c$ to adequately remove any non-zero magnetic field divergence. The definition of $\mathbf{J}$, the electric current, will be discussed later. The fluid equations follow. The continuity equations for ions (subscript "$i$") and neutrals (subscript "$n$") are given as

\begin{equation}
\frac{\partial n_{i}}{\partial t}+\nabla \cdot\left(n_{i} \mathbf{V}_{i}\right)=\Gamma_{i}^{\mathrm{ion}}+\Gamma_{i}^{\mathrm{rec}}, \quad \text{and}
\end{equation}
\begin{equation}
\frac{\partial n_{n}}{\partial t}+\nabla \cdot\left(n_{n} \mathbf{V}_{n}\right)=\Gamma_{n}^{\mathrm{rec}}+\Gamma_{n}^{\mathrm{ion}},
\end{equation}

\noindent  respectively, where the $\Gamma$ terms, the source rates, on the right-hand sides correspond to ionisation and recombination rate parameters, specifically

\begin{equation}
\Gamma_{i}^{\mathrm{rec}} \equiv-n_{i} v^{\mathrm{rec}} \qquad \text{and} \qquad   \Gamma_{n}^{\mathrm{ion}} \equiv-n_{n} v^{\mathrm{ion}},
\end{equation}

\noindent with

\begin{equation}
      v^{\mathrm{rec}}=n_{e} \frac{1}{\sqrt{T_{e}^{*}}} 2.6 \times 10^{-19} \mathrm{~m}^{3} \mathrm{~s}^{-1}
\end{equation}

\noindent and 

\begin{equation}
v^{\mathrm{ion}}=n_{e} A \frac{1}{X+\phi_{\mathrm{ion}} / T_{e}^{*}}\left(\frac{\phi_{\mathrm{ion}}}{T_{e}^{*}}\right)^{K} e^{-\phi_{\text {ion }} / T_{e}^{*}} \mathrm{~m}^{3} \mathrm{~s}^{-1}.
\end{equation}

In the above expressions, the ionisation potential, $\phi_{\text {ion }}$, is set to $ 13.6$ $\mathrm{eV}$ (i.e. the ionisation potential of the hydrogen atom) and the constants are given as

\begin{equation}
A=2.91 \times 10^{-14}, \quad K=0.39, \quad \text{and} \quad X=0.232,
\end{equation}

\noindent  according to \citet{MeierShumlak2012}. {While in \citet{Leake2012} these values were used only for a weakly or partially ionised plasma (hence, not the corona), as seen in \citet{Voronov1997}, the ionisation rates stated above are valid up to 20~keV. Though this is not the case of the recombination frequency, $\nu^\text{rec}$, from Figure~3 of the analysis of \cite{Nahar2021}, it can be seen that the resulting recombination cross-section is also approximately correct for temperatures of the order of a million of Kelvin.} Next, the momentum equation for the ions is

\begin{eqnarray}
\nonumber
\frac{\partial}{\partial t}\left(m_{i} n_{i} \mathbf{V}_{i}\right)+\nabla \cdot\left(m_{i} n_{i} \mathbf{V}_{i} \mathbf{V}_{i}+\mathbb{P}_{i}\right)
=q_{i} n_{i}\left(\mathbf{E}+\mathbf{V}_{i} \times \mathbf{B}\right)+ \\ \mathbf{R}_{i}^{i n}+ \mathbf{S}^\mathrm{ion}_i + \mathbf{S}^\mathrm{cx}_i + m_i n_i \mathbf{g},
\end{eqnarray}

\noindent in which the terms related to the momentum transfer due to ionisation and charge exchange are defined as

\begin{equation}
     \mathbf{S}^\mathrm{ion}_i = \Gamma_{i}^\mathrm{ion} m_{i} \mathbf{V}_{n} -\Gamma_{n}^\mathrm{rec} m_{i} \mathbf{V}_{i}  
\end{equation}
and
\begin{equation}
\label{eq:scxn}
     \mathbf{S}^\mathrm{cx}_i = \Gamma^\mathrm{cx} m_{i}\left(\mathbf{V}_{n}-\mathbf{V}_{i}\right)+\mathbf{R}_{i n}^\mathrm{cx}-\mathbf{R}_{n i}^\mathrm{cx}.
\end{equation}

\noindent The momentum equation for the neutrals is

\begin{eqnarray}
\nonumber
\frac{\partial}{\partial t}\left(m_{n} n_{n} \mathbf{V}_{n}\right)+\nabla \cdot\left(m_{n} n_{n} \mathbf{V}_{n} \mathbf{V}_{n}+\mathbb{P}_{n}\right)
=-\mathbf{R}_{i}^{i n} + \mathbf{S}^\mathrm{ion}_n + \\ \mathbf{S}^\mathrm{cx}_n + m_n n_n \mathbf{g},
\end{eqnarray}

\noindent  where the respective ionisation and charge exchange terms are given by

\begin{equation}
    \mathbf{S}^\mathrm{ion}_n = \Gamma_{n}^\mathrm{rec} m_{i} \mathbf{V}_{i} -\Gamma_{i}^\mathrm{ion} m_{n} \mathbf{V}_{n}  
\end{equation}
and
\begin{equation}
\label{eq:scx}
    \mathbf{S}^\mathrm{cx}_n = \Gamma^{c x} m_{i}\left(\mathbf{V}_{i}-\mathbf{V}_{n}\right)-\mathbf{R}_{i n}^\mathrm{cx}+\mathbf{R}_{n i}^\mathrm{cx}.
\end{equation}

\noindent In the expressions above, the collisional momentum transfer between the species is defined as

\begin{equation}
\mathbf{R}_{i}^{in}=m_{i n} n_{i} \nu_{i n}\left(\mathbf{V}_{n}-\mathbf{V}_{i}\right) \quad \text{and} \quad  \mathbf{R}_{n}^{n i}=m_{n i} n_{n} \nu_{n i}\left(\mathbf{V}_{i}-\mathbf{V}_{n}\right),
\end{equation}

\noindent with the collisional frequencies, $ \nu_{i n}$ and $ \nu_{n i}$, of

\begin{equation}
\nu_{i n}=n_{n} \Sigma_{i n} \sqrt{\frac{8 k_{B} T_{i n}}{\pi m_{i n}}} \quad \text{and} \quad \nu_{n i}=n_{i} \Sigma_{n i} \sqrt{\frac{8 k_{B} T_{n i}}{\pi m_{n i}}}. 
\label{eq:collisionalrate}
\end{equation}

\noindent Here, $m_{i n}$, $m_{n i}$ and $T_{i n}$, $T_{n i}$ are defined as

\begin{equation}
m_{i n}=\frac{m_{i} m_{n}}{m_{i}+m_{n}} = m_{n i} \quad  \text{and} \quad T_{i n}=\frac{T_{i}+T_{n}}{2}= T_{n i}.
\end{equation}

\noindent Furthermore, the collisional cross-sections, $\Sigma$, for ions and neutrals are given by

\begin{equation}
\label{eq:chargeexchangesigma}
\Sigma_{\mathrm{ni}}=\Sigma_{\mathrm{in}} = 1.41\cdot10^{-19} \text{m}^2,
\end{equation}
respectively. {Just like above, these values were originally used only for a partially or weakly ionised plasma. If we consult Figure 7 of the study of \cite{Wargnier_2022}, however, we find that the above presented cross-section value is actually within the range of values that would be expected.}

Finally, the terms describing the momentum transfer due to the charge exchange in Eq.~\eqref{eq:scxn} are given by

\begin{equation}
\mathbf{R}_{\mathrm{in}}^{\mathrm{cx}} \approx-m_{i} \sigma_{\mathrm{cx}}\left(V_{\mathrm{cx}}\right) n_{i} n_{n} \mathbf{V}_{\mathrm{in}} v_{T n}^{2}\left[4\left(\frac{4}{\pi} v_{T i}^{2}+v_{\mathrm{in}}^{2}\right)+\frac{9 \pi}{4} v_{T n}^{2}\right]^{-1 / 2}
\end{equation}

\noindent and

\begin{equation}
\mathbf{R}_{\mathrm{ni}}^{\mathrm{cx}} \approx m_{i} \sigma_{\mathrm{cx}}\left(V_{\mathrm{cx}}\right) n_{i} n_{n} \mathbf{V}_{\text {in }} v_{T i}^{2}\left[4\left(\frac{4}{\pi} v_{T n}^{2}+v_{\mathrm{in}}^{2}\right)+\frac{9 \pi}{4} v_{T i}^{2}\right]^{-1 / 2},
\end{equation}

\noindent respectively, where $v_{T \alpha}=\sqrt{2 k_{B} T_{\alpha} / m_{\alpha}}$ is the species thermal velocity and

\begin{equation}
V_{c x} \equiv \sqrt{\frac{4}{\pi} v_{T i}^{2}+\frac{4}{\pi} v_{T n}^{2}+v_{i n}^{2}} \quad \text{and} \quad v_{i n}^{2} \equiv\left|\mathbf{V}_{i}-\mathbf{V}_{n}\right|^{2},
\end{equation}

\noindent with $V_{c x}$ being the representative speed of the charge exchange interaction and $v_{i n}$ the relative speed of neutrals with respect to ions. 

The cross-section for charge exchange for hydrogen can be approximated by

\begin{equation}
\sigma_{c x, H}=1.09 \times 10^{-18}-7.15 \times 10^{-20} \ln \left(V_{c x}\right) ~\mathrm{m}^{2},
\end{equation}

\noindent from which the $\Gamma^{c x}$ term from Equation \eqref{eq:scx} was also determined as

\begin{equation}
\Gamma^{c x} \equiv \sigma_{c x}\left(V_{c x}\right) n_{i} n_{n} V_{c x},
\end{equation}

{where the charge exchange cross-section, $\sigma_{cx}$, is valid over the range of temperatures from 0.12 eV to 10 keV \citep{MeierShumlak2012}.}

Finally, the energy equation for the ions is given by

\begin{eqnarray}
\nonumber
\frac{\partial \varepsilon_{i}}{\partial t}+\nabla \cdot\left(\varepsilon_{i} \mathbf{V}_{i}+\mathbf{V}_{i} \cdot \mathbb{P}_{i}+\mathbf{h}_{i}\right)
=\mathbf{V}_{i} \cdot\left(q_{i} n_{i} \mathbf{E}+ \mathbf{R}_{i}^{i n} + m_i n_i \mathbf{g} \right)+ \\ Q_{i}^{i n} + S^\mathrm{ion}_i + S^\mathrm{cx}_i,
\end{eqnarray}

\noindent where the ionisation and charge-exchange-related energy transfer ($S$-) terms are given by

\begin{equation}
     S^\mathrm{ion}_i = \frac{m_{i}}{m_{n}}\left(\Gamma_{i}^\mathrm{ion} \frac{1}{2} m_{n} V_{n}^{2} + Q_{n}^\mathrm{ion}\right) - \Gamma_{n}^\mathrm{rec} \frac{1}{2} m_{i} V_{i}^{2} - Q_{i}^\mathrm{rec}  
\end{equation}
and
\begin{equation}
     S^\mathrm{cx}_i = \Gamma^\mathrm{cx} \frac{1}{2} m_{i}\left(V_{n}^{2} - V_{i}^{2}\right) + \mathbf{V}_{n} \cdot \mathbf{R}_{i n}^\mathrm{cx} - \mathbf{V}_{i} \cdot \mathbf{R}_{n i}^\mathrm{cx} + Q_{i n}^\mathrm{cx} - Q_{n i}^\mathrm{cx},
\end{equation}
respectively.

\noindent The energy equation for the neutrals is

\begin{eqnarray}
\nonumber
\frac{\partial \varepsilon_{n}}{\partial t}+\nabla \cdot\left(\varepsilon_{n} \mathbf{V}_{n}+\mathbf{V}_{n} \cdot \mathbb{P}_{n}+\mathbf{h}_{n} \right)=-\mathbf{V}_{n} \cdot\left(\mathbf{R}_{i}^{i n}+ m_n n_n \mathbf{g}\right) + \\ Q_{n}^{n i} + S^\mathrm{ion}_n + S^\mathrm{cx}_n,
\end{eqnarray}

\noindent where, now, the ionisation and charge-exchange-related energy transfer terms are defined as
\begin{equation}
     S^\mathrm{ion}_n = \Gamma_{n}^\mathrm{rec}  \frac{1}{2} m_{i} V_{i}^{2}  + Q_{i}^\mathrm{rec} - \Gamma_{i}^\mathrm{ion}  \frac{1}{2} m_{n} V_{n}^{2} - Q_{n}^\mathrm{ion}  
\end{equation}
and
\begin{equation}
     S^\mathrm{cx}_n = \Gamma^\mathrm{cx} \frac{1}{2} m_{i}\left(V_{i}^{2}-V_{n}^{2}\right) + \mathbf{V}_{i} \cdot \mathbf{R}_{n i}^\mathrm{cx} - \mathbf{V}_{n} \cdot \mathbf{R}_{i n}^\mathrm{cx} + Q_{n i}^\mathrm{cx} - Q_{i n}^\mathrm{cx},
\end{equation}

\noindent respectively. Here, we used the definitions 

\begin{equation}
Q_{n}^\mathrm{ion} \equiv \Gamma_{i}^\mathrm{ion} \frac{3}{2} k T_{n} \quad \text{and} \quad  Q_{i}^\mathrm{rec} \equiv \Gamma_{n}^\mathrm{rec} \frac{3}{2} k T_{i}.
\label{eq:no33}
\end{equation}

The collisional energy transfer terms, $Q_i^{in}$, $Q_n^{ni}$, and the charge exchange energy transfer terms, $Q_{i n}^{c x}$, $Q_{n i}^{c x}$, are defined as (see \cite{Leake2012} and \cite{MeierShumlak2012}):

\begin{equation}
    Q_i^{in} = \mathbf{R}_i^{in} \cdot (\mathbf{V}_n - \mathbf{V}_i) + m_{in} n_i \nu_{in} (T_n - T_i), 
\end{equation}

\begin{equation}
    Q_n^{ni} = \mathbf{R}_n^{ni} \cdot (\mathbf{V}_i - \mathbf{V}_n) + m_{ni} n_n \nu_{ni} (T_i - T_n),
\end{equation}

\begin{equation}
Q_{i n}^{c x} \approx \sigma_{c x}\left(V_{c x}\right) m_i n_i n_n \frac{3}{4} v_{T n}^2 \sqrt{\frac{4}{\pi} v_{T i}^2+\frac{64}{9 \pi} v_{T n}^2+v_{i n}^2}, \quad \text{and}
\end{equation}

\begin{equation}
Q_{n i}^{c x} \approx \sigma_{c x}\left(V_{c x}\right) m_i n_i n_n \frac{3}{4} v_{T i}^2 \sqrt{\frac{4}{\pi} v_{T n}^2+\frac{64}{9 \pi} v_{T i}^2+v_{i n}^2} .
\label{eq:nolastresistive}
\end{equation}

In order to close the set of fluid equations and to couple them back to the Maxwell equations, the form of the generalised Ohm's formula from the PhD thesis of \citet{jones2011} is used as an approximation for the computation of the electric current density. The magnitude of the latter is adjusted to ensure the stability of the code while keeping it as low as possible, similar to how it is implemented in the code Bifrost according to  \cite{Charalambos2019}.

It should be noted that the ion fluid described above is assumed to contain the electron species in it in order to create a charge-neutral system. Thus, it accounts for the electron pressure and electron charge (under the assumption of quasi-neutrality) but neglects the electron mass. {How the electrons are accounted for in the equations, along with the lack of radiative losses and the inclusion of gravity, are the only differences between the energy equation presented here (and used in our simulations) when compared to \cite{Leake2012}.}

A certain portion of this paper is devoted to a comparison of the results with the default COCONUT version based on ideal MHD; hence, the MHD formulation of the default COCONUT set-up is given below for convenience:

\begin{equation}
    \frac{d\rho}{dt} + \mathbf{\nabla} \cdot (\rho \mathbf{V}) = 0,
\label{eq:no34}
\end{equation}

\begin{equation}
    \frac{d(\rho \mathbf{V})}{dt} + \mathbf{\nabla} \cdot \left( \rho \mathbf{V} \otimes \mathbf{V} + \mathbf{I} \left( P + \frac{1}{2}|\mathbf{B}|^2 \right) - \mathbf{B} \otimes \mathbf{B}  \right) = \rho \mathbf{g},
\end{equation}

\begin{equation}
    \frac{d\varepsilon}{dt} + \mathbf{\nabla} \cdot \left( \left( \varepsilon + P + \frac{1}{2}|\mathbf{B}|^2 \right)  \mathbf{V} - \mathbf{B} ( \mathbf{V} \cdot \mathbf{B})  \right) = \rho \mathbf{g} \cdot \mathbf{V},
\end{equation}

\begin{equation}
    \frac{d\mathbf{B}}{dt}  +  \mathbf{\nabla} \cdot \left( \mathbf{V} \otimes \mathbf{B} -  \mathbf{B}  \otimes \mathbf{V} + \mathbf{I} \phi \right) = \mathbf{0},
\label{eq:E10}
\quad \text{and}
\end{equation}

\begin{equation}
    \frac{d\phi}{dt} + \mathbf{\nabla} \cdot \left( V^2_\text{ref} \mathbf{B} \right) = 0,
\label{eq:no38}
\end{equation}

\noindent where $\varepsilon$ is the non-dimensional internal energy, $\phi$ the divergence cleaning parameter and $\mathbf{I}$ the identity dyadic. {While in the MFMHD set-up we use a realistic $\gamma = 5/3$, for reasons already outlined in the original publication \citepalias{PerriLeitner2022} the default ideal MHD set-up assumes $\gamma \approx 1$. }

{Thus, Equations \ref{eq:no1} through \ref{eq:nolastresistive} are implemented in the resistive MFMHD set-up and Equations \ref{eq:no34} through \ref{eq:no38} by the ideal MHD (single-fluid) set-up.}

The clearest difference between the two formulations is the fact that the MFMHD set-up is formulated for ions and neutrals separately. Beyond this, however, there are also differences in the physics that is included, namely:
\begin{itemize}
    \item the MFMHD formulation assumes resistivity in Ohm's law and considers the electric field as a primitive variable, and
    \item the MFMHD formulation has additional mass, momentum, and energy sources related to collisions, ionisation, recombination, and charge exchange.
\end{itemize}

In addition, due to the fact that the original MFMHD set-up in COOLFluiD has been developed using a different scheme than the ideal MHD set-up (see e.g. \citet{ALVAREZLAGUNA2016252} and \citet{Laguna2018}), the set-ups numerically further vary as follows:

\begin{itemize}
    \item while the MFMHD approach uses the AUSM+up (Advection Upstream Splitting Method) scheme for discretizing the convective fluxes of the fluid equations, the MHD module uses HLL (Harten-Lax-van Leer), which is more dissipative and less accurate,
    \item while the MFMHD module solves the equations in a dimensional form, the MHD module solves them non-dimensionally,
    \item since MFMHD solves for both $\mathbf{B}$ and $\mathbf{E}$, it requires two cleaning parameters in the hyperbolic divergence cleaning method compared to just one (for $\mathbf{B}$) in MHD, and
    \item since the AUSM+up scheme is less dissipative than HLL, it requires more limiting, and thus the Venkatakrishnan limiting coefficients differ (see \citet{Venkatakrishnan1993}). 
\end{itemize}

Below, other critical aspects of the two simulation set-ups are discussed and compared. 

\subsection{The grid}

   \begin{figure*}
   \centering
   \includegraphics[width=15cm]{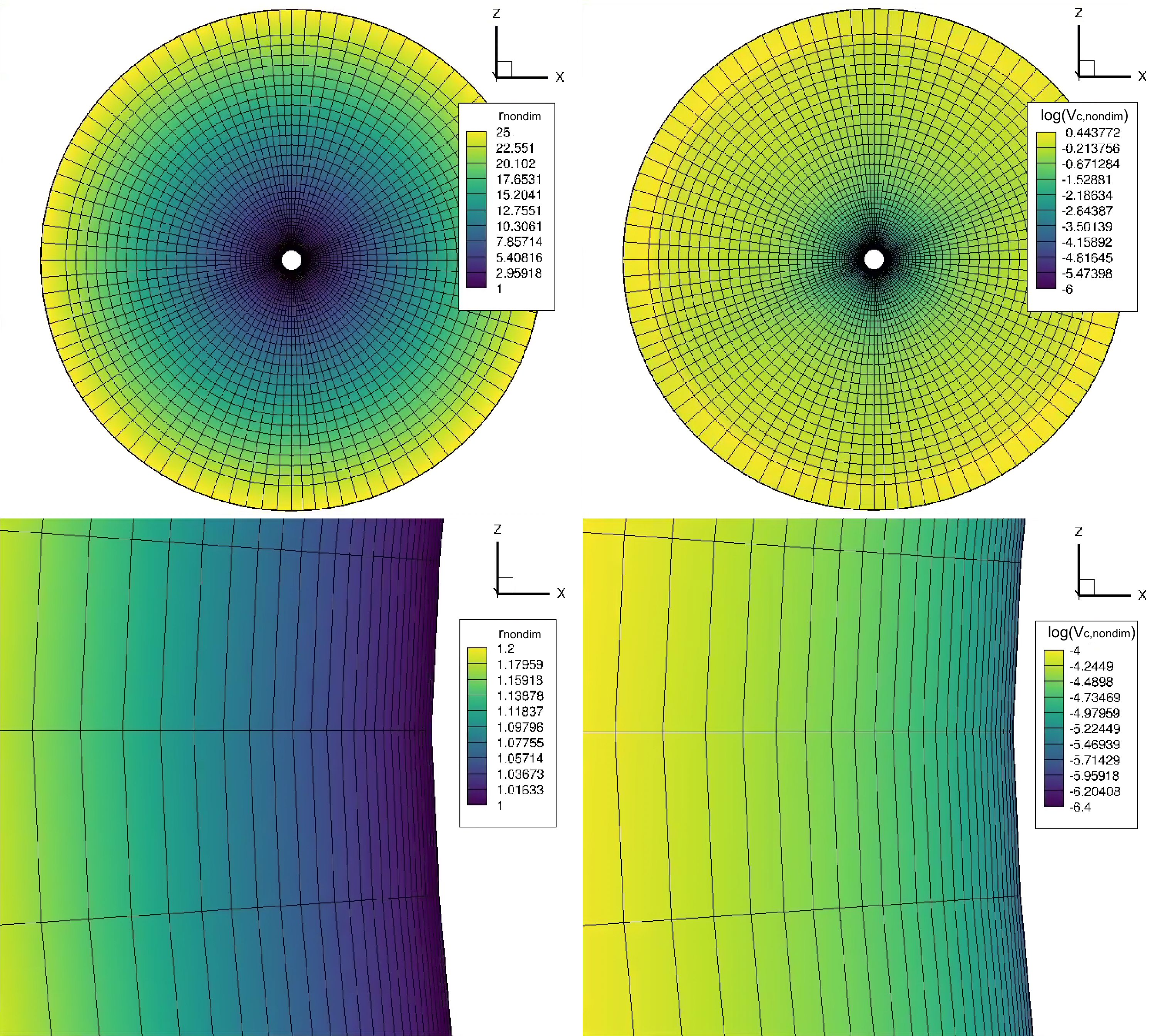}
   \caption{Radial distribution of the grid cells for the whole domain (top left) and near the inner boundary (bottom left) in $\;R_\odot$ and the corresponding logarithm of the cell volume in $\;R_\odot^3$ (right).}
              \label{fig:grid}%
    \end{figure*}
    
The grid that was used for both set-ups (MHD and MFMHD) corresponds to the standard grid that is used with the COCONUT solver (see \citet{Brchnelova2022a}), that is to say, unstructured (based on a subdivided icosahedron radially extended outward) spanning from 1.01$\;R_\odot$ to 25 $R_\odot$. 
{In Fig.~\ref{fig:grid}, the top row shows the full domain, with the 1.01$\;R_\odot$ boundary in the middle (representing the lower corona) and  25 $R_\odot$ on the outside. The left-hand side shows the radial distance from the centre of the Sun whereas the right side shows the logarithm of the cell volume, in the units of $R_\odot^3$. In the bottom row, there is a close-up of the left side of the inner boundary. It can be seen that the smallest boundary cells have a volume of roughly $10^{-6.5} R_\odot^3$. }

{The current grid can just resolve the smallest collisional length scales near the inner boundary. The neutral-ion collisional length scale near the inner boundary, computed from the collisional rates defined in Equation \ref{eq:collisionalrate}, is of the order of $10^5$m. The width of our smallest cell is $2.5 \cdot 10^{-4}  R_\odot$, which is thus also of $10^{5}$m. Further away from the boundary, the density decreases significantly and the collisional length scales increase accordingly. After a drop in density of three to four magnitudes, the limiting collisional length scale becomes larger than 1 $R_\odot$, while the width of our largest cells remains constrained to a maximum of $R_\odot$ near the outer boundary.}

{In contrast to the collisional length scales, the current grid is not intended to be fine enough to resolve very local ion-neutral phenomena such as ion-neutral waves. However, waves and other unsteady phenomena are not the focus of the present study anyway, since the solver is currently steady-state. For space weather applications in the corona the structures that are the most relevant for the accuracy of the solar wind are for example streamers, loops, and coronal holes, for which the general resolution and distribution of the current grid was found to be sufficient (see e.g. \cite{Kuzma2023}).} 

The number of grid cells for each simulation was 300k (unless stated otherwise for some of the ideal MHD simulations) since, as is shown later, this was sufficient to resolve the important features in the MFMHD flow field with the less diffusive AUSM+up scheme. 

\subsection{Initial \& boundary conditions}
 
The initial conditions and the boundary conditions for the MHD simulations are described in detail in \citet{Brchnelova2022b}. The same boundary conditions are used for ions in the MFMHD simulations, with the only difference being that the temperature is treated as a primitive variable instead of pressure. Thus, the boundary ion temperature was prescribed to be the same as the resulting temperature computed from the boundary density and boundary pressure in the ideal MHD set-up. 

For the neutrals, we prescribe a $10^{-6}$ concentration at the inner boundary (an estimate of the concentration given the inner boundary temperature) and the same temperature as that of the ions. As for the initial conditions, due to the fact that the MFMHD simulations are far less robust and take longer to compute, they are generally initiated from the post-processed MHD solution as it is closer to the final solution than any other initial guess we could provide. For the neutrals, their initial concentration is assumed to be $10^{-6}$ in the entire domain, with the same temperature as the ions and zero velocity. Naturally, the neutral fraction will change in the domain as the simulation converges, due to the ionisation and recombination terms. This was tested by starting the simulation at a higher neutral density, which still converged to the same solution.

   \begin{figure*}
   \centering
   \includegraphics[width=15cm]{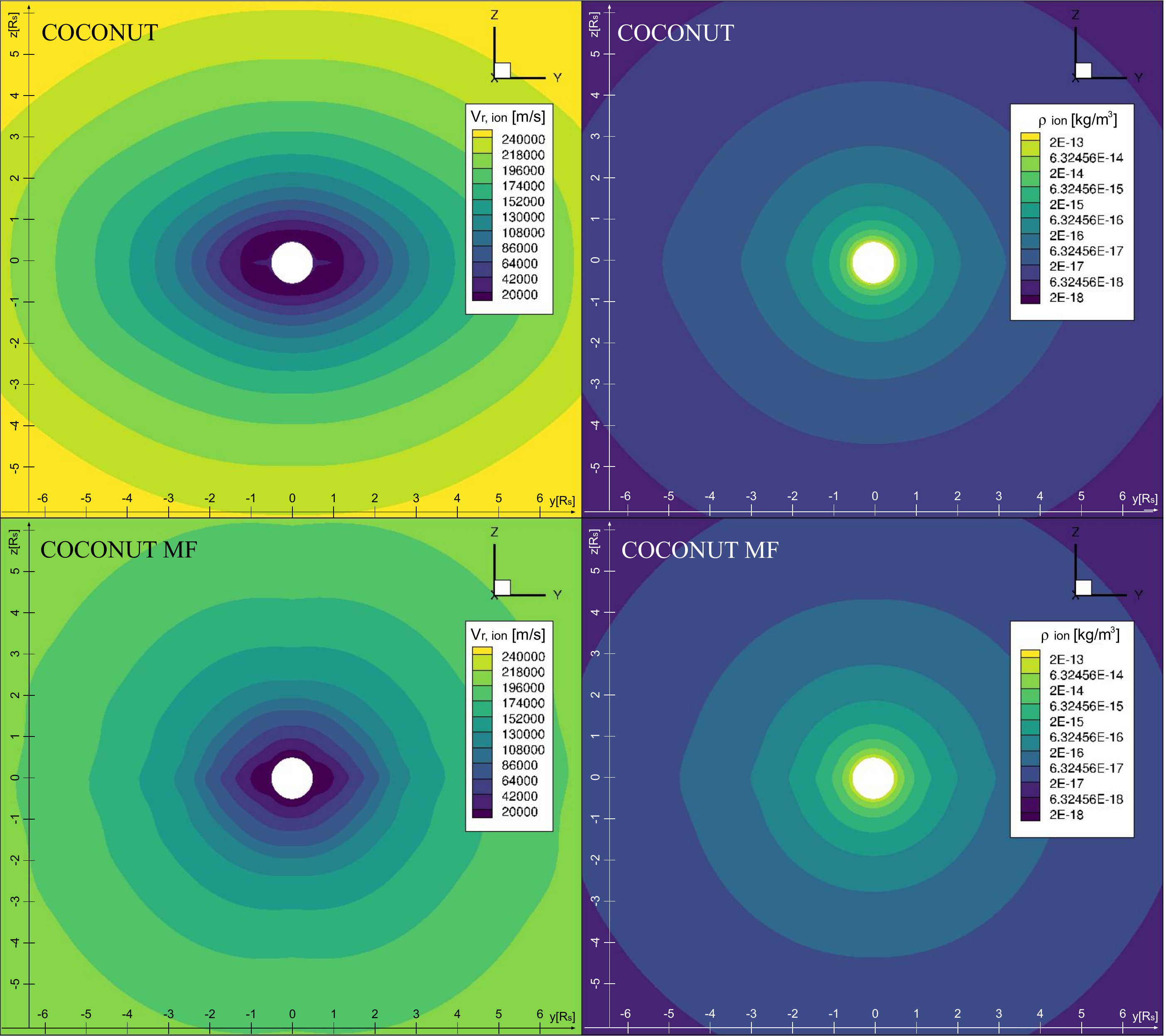}
   \caption{Comparison between the ideal MHD and MFMHD simulation results for a simple magnetic dipole. The top row shows the ideal MHD case at the same grid resolution as the MFMHD case in the second row. The left panels show the ion radial velocity and the right panels show the ion density profiles.}
              \label{fig:dipolecomp}%
    \end{figure*}

\subsection{Limitations}

It should be noted that even though this set-up was developed to determine the effects of resolving the ion and neutral fluids separately on the resolved global coronal dynamics, its formulation and resolution do prevent it from studying some of these phenomena in more detail. For example, since our set-ups (both the ideal MHD and the MFMHD) are steady-state, ion-neutral waves are not modelled in a time-accurate fashion. But even if a time-accurate scheme were implemented, since global corona models require relatively large cell sizes to cover the entire domain, these waves would still not be well resolved. Thus, the results of this study do not consider possible wave effects. 

In addition, here we only experiment with a well-tested ideal MHD polytropic set-up, and thus an equivalent polytropic MF set-up. In these simulations, we do not include approximations of coronal heating mechanisms, radiation, and thermal conduction, which means that the resolved solar wind is not bimodal. Work is currently ongoing to include all these terms in the baseline MHD COCONUT model. Once this is achieved, verified, and validated, these terms will be included in the MF simulations as well. 


\section{Application to a magnetic dipole}
\label{sec:dipole}

Firstly, the new formulation was applied to a magnetic dipole for validation and comparison with the result of the ideal MHD model. Apart from the differences in the set-ups outlined above, the test cases were identical. The results are shown in Fig.~\ref{fig:dipolecomp}, where the bottom row of images corresponds to the MFMHD version and the top row of images corresponds to the baseline ideal MHD COCONUT setting. 

It is clear from Fig.~\ref{fig:dipolecomp} that a dipolar structure of a similar shape is obtained in both cases and also that the value ranges for both $V_r$ and $\rho$ are similar (note that the colour bar ranges are chosen to be the same). However, the exact shape of the structures, the resulting velocities, and densities vary in a visible way. While some of the differences may be explained by the different schemes and the amount of limiting applied, a significant contribution to these differences comes from the collisional and resistive terms that the ideal MHD model neglects. Interactions of the ions with the neutral particles and the resulting resistive terms (present both in the source terms in the Euler equations and in the generalised Ohm's law) will generally dissipate some of the momentum. That is what can be seen, especially in the two left-hand side panels of Figure~\ref{fig:dipolecomp}, with the MHD simulation result reaching a higher velocity. When it comes to maximum values, the differences in the total density are only very small since the concentration of neutrals is much smaller than that of ions. The variation in shape of the isosurfaces in this case can be attributed to the different numerical dissipation in the domain, due to the different numerical scheme, as well as to the use of the resistive Ohm's law in the MFMHD model.

\section{Application to real (data-driven) cases}
\label{sec:applicationreal}

   \begin{figure*}
   \centering
   \includegraphics[width=19cm]{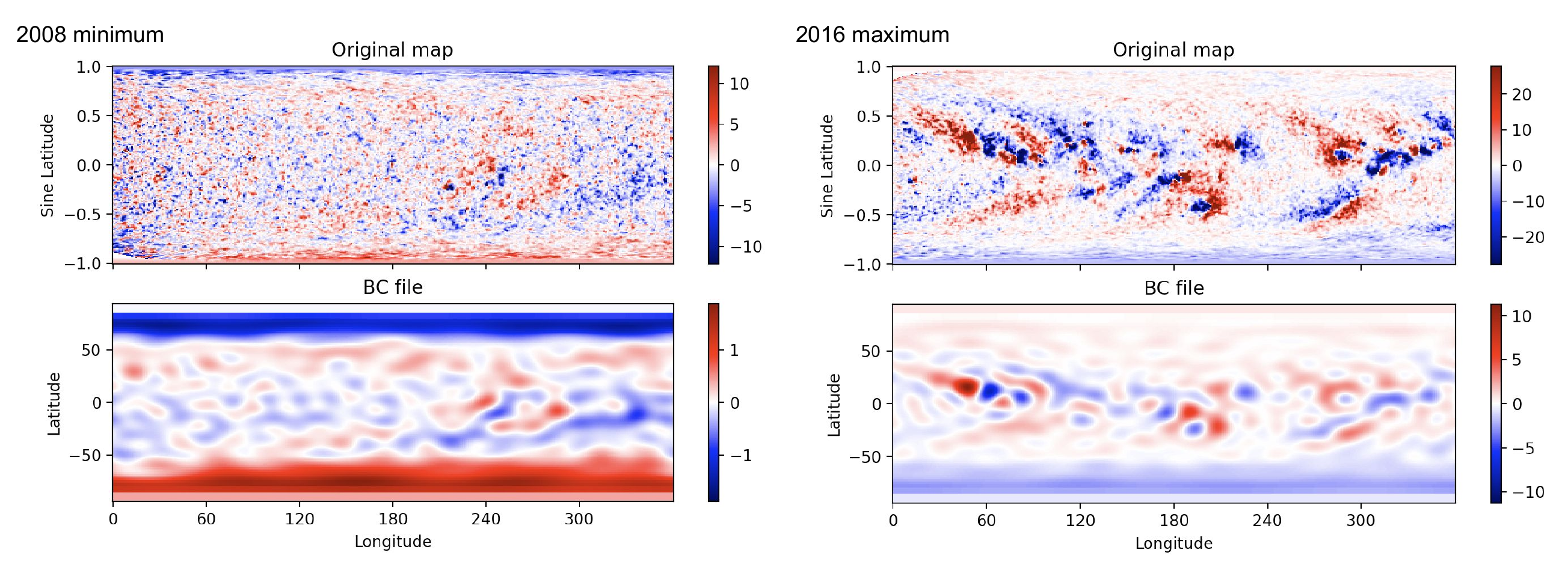}
   \caption{Original photospheric magnetograms (top row) for the 2008 (left) and 2016 (right) solar eclipses and derived magnetic maps (bottom row) prescribed at the inner boundary after spherical harmonics processing.}
              \label{fig:raw_maps}%
    \end{figure*}

   \begin{figure*}
   \centering
   \includegraphics[width=17cm]{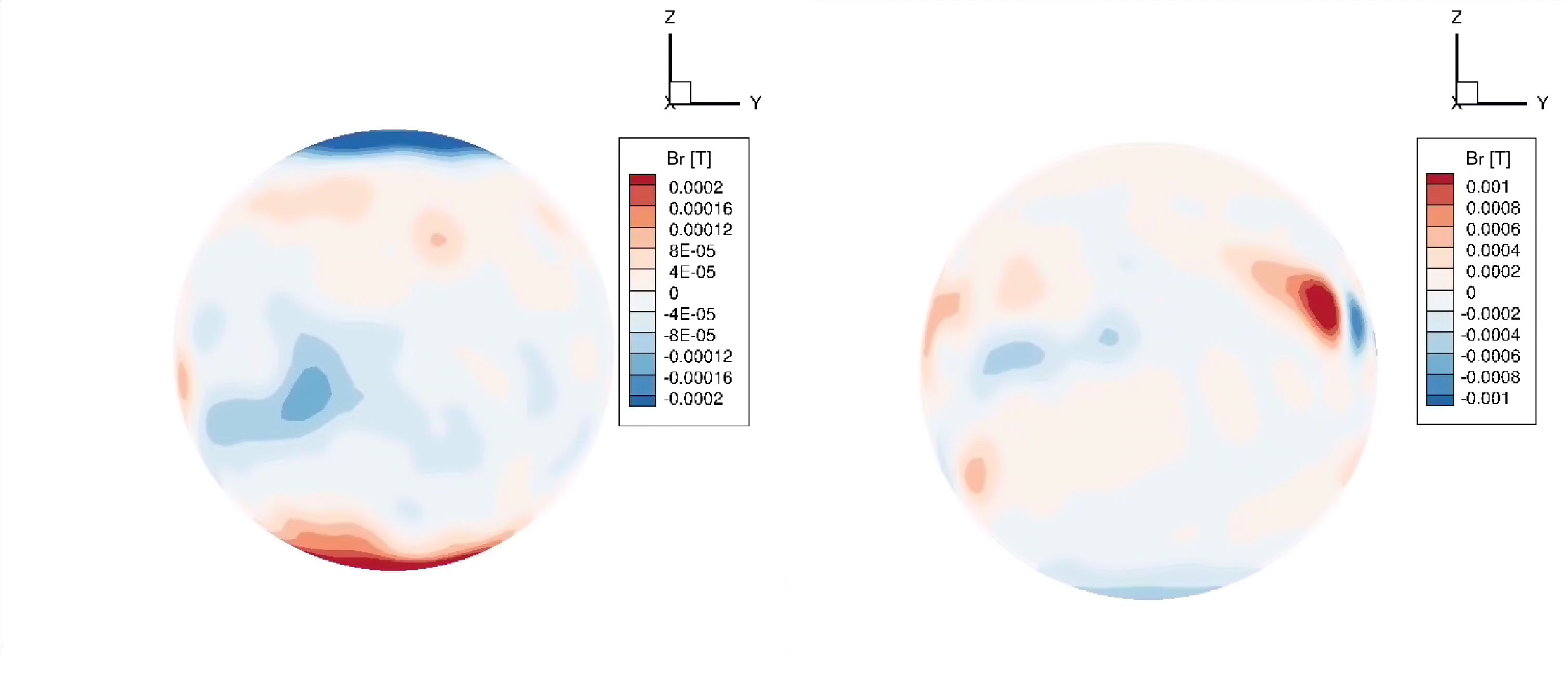}
   \caption{Boundary magnetic maps at the basic orientation (0 degree longitude) of the two cases simulated, August 1, 2008 (solar minimum) on the left and March 9, 2016 (solar maximum) on the right, using the prescription according to the bottom maps shown in Figure \ref{fig:raw_maps}.}
              \label{fig:maps}%
    \end{figure*}
    
Next, we test the solver on data-driven simulations, using a photospheric magnetogram as input for the magnetic field at the lower boundary, and evaluate the effects of the interactions of the ions with the neutrals on the plasma dynamics. The simulations were performed with minimum and maximum solar activity GONG (Global Oscillation Network Group) magnetograms, which corresponded to the August 1, 2008 eclipse and the March 9, 2016 eclipse, respectively. These dates were chosen since a validation was already performed for the ideal MHD COCONUT model for these cases (see e.g. \citet{Kuźma23}). {The selected magnetograms are shown in Fig.~\ref{fig:raw_maps}, where on the top row the photospheric magnetograms are shown in Gauss for both cases. In the bottom row, it is shown what these magnetic maps look like after the smoothing process through spherical harmonics (since the magnetic field gradients and strengths are expected to be much lower in the lower corona compared to the photosphere from where the magnetic maps are derived).} For a discussion about the required pre-processing and the possible types of magnetograms that can be used for COCONUT simulations, consult \citet{Perri2022a} or \cite{Brchnelova2023}.

The magnetograms at the default orientation (0 degree latitude) are shown in Fig.~\ref{fig:maps}, with the minimum case on the left and the maximum case on the right. The solar minimum magnetogram has maximum magnetic field values of 2-3 G, whereas the solar maximum magnetogram reaches 10-12 G in active regions. 

\subsection{Comparison with the ideal MHD results}

\begin{figure*}
   \centering
   \includegraphics[width=15cm]{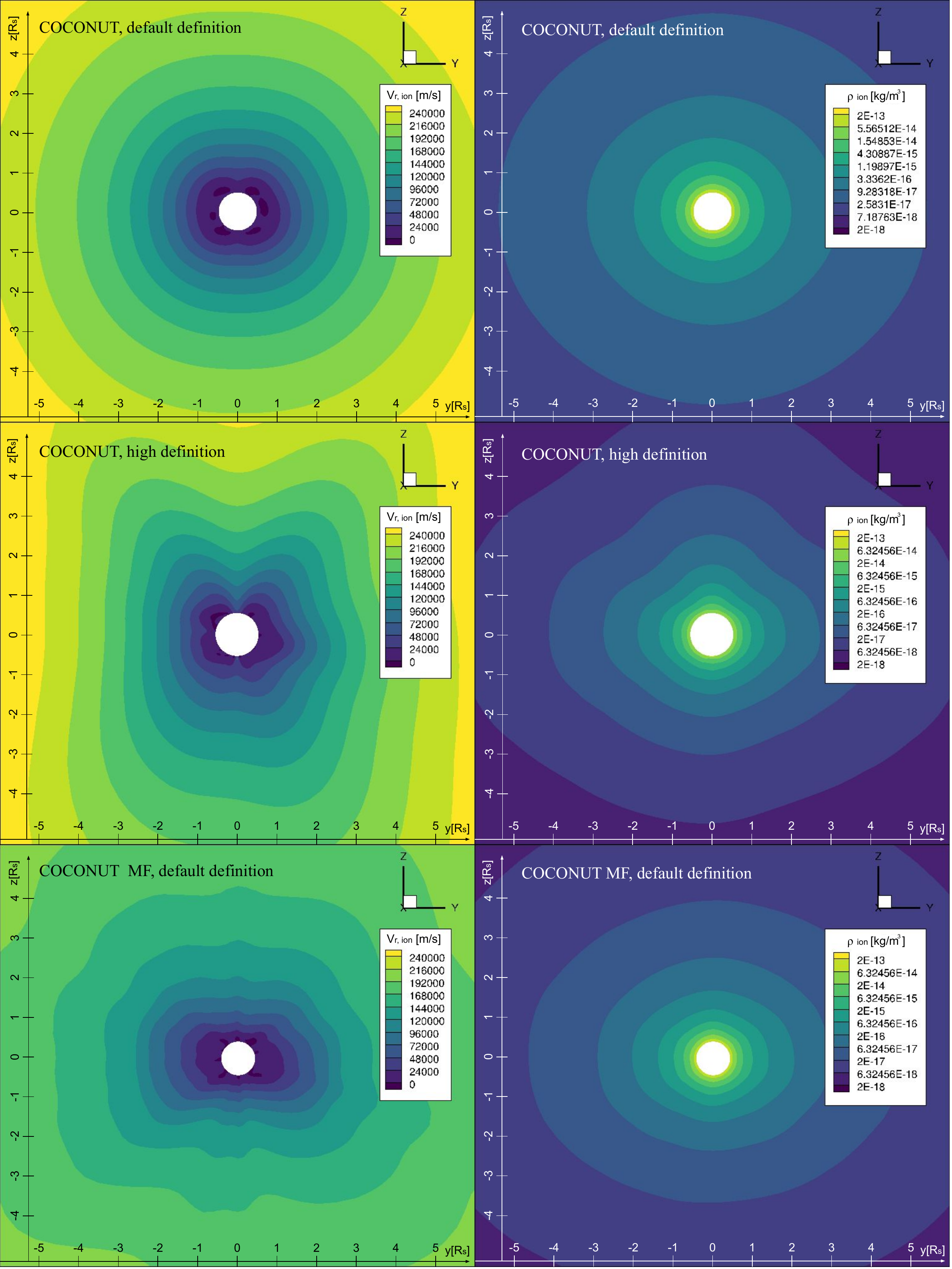}
   \caption{Comparison between the ideal MHD and MFMHD simulations of the solar minimum case. The top row shows the ideal MHD case at the same grid resolution as the MFMHD case in the bottom row. Since this solution is much more dissipated due to the different scheme, a higher resolution, lower-dissipation ideal MHD solution is shown in the middle to confirm the correct positioning of the streamers resolved in the MFMHD result. The left panels show the ion radial velocity and the right panels the ion density profiles.}
              \label{fig:Minimum_mhdvrrho_mfmhdvrrho}%
    \end{figure*}

\begin{figure*}
   \centering
   \includegraphics[width=15cm]{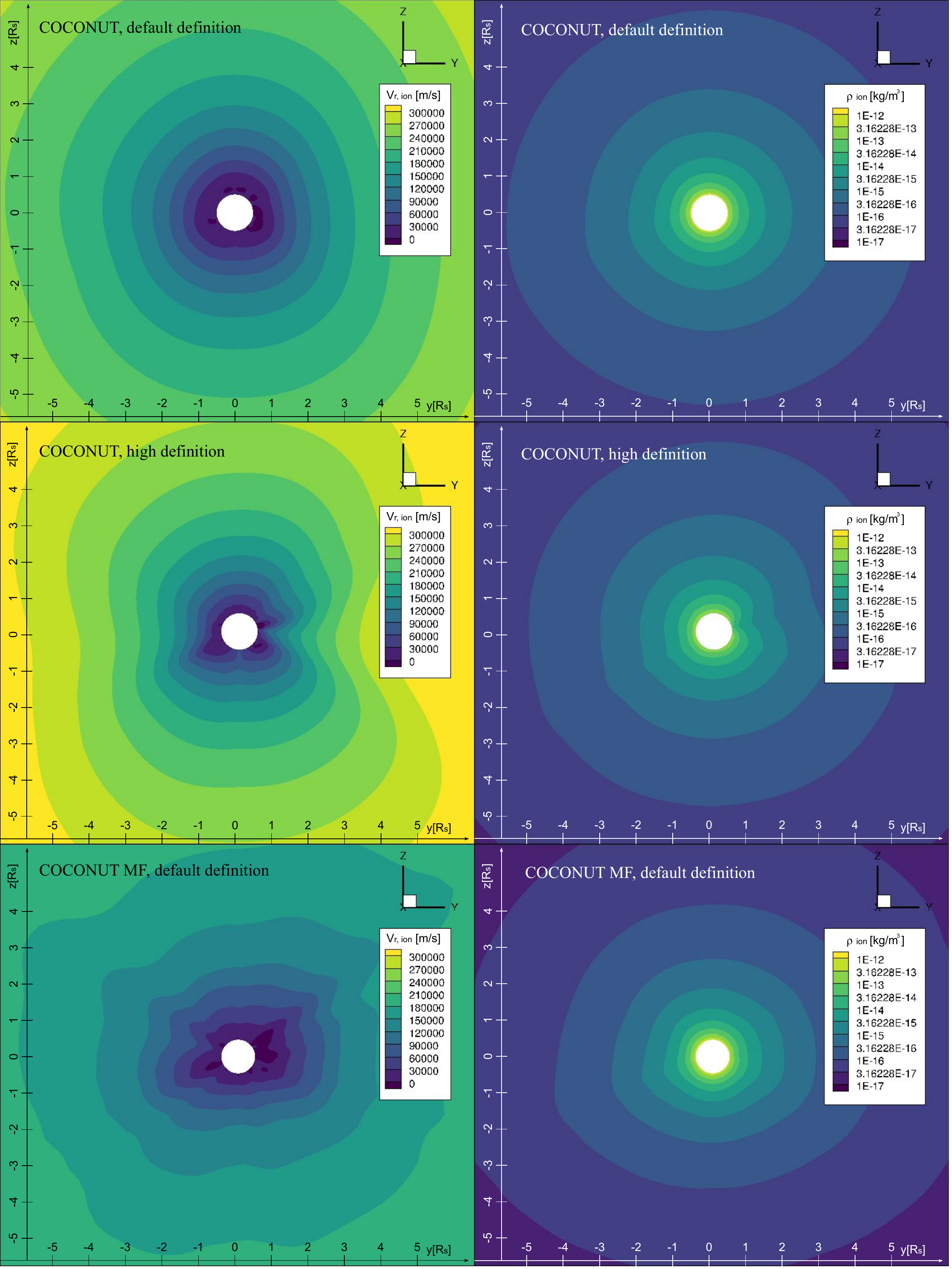}
   \caption{Comparison between the ideal MHD and MFMHD simulations of the solar maximum case. The top row shows the ideal MHD case at the same grid resolution as the MFMHD case in the bottom row. Since this solution is much more dissipated due to the different scheme, a higher resolution, lower-dissipation ideal MHD solution is shown in the middle to confirm the correct positioning of the streamers resolved in the MFMHD result. The left panels show the ion radial velocity and the right panels the ion density profiles.}
              \label{fig:Maximum_mhdvrrho_mfmhdvrrho}%
    \end{figure*}

Let us compare the results of the MFMHD model with those of the basic MHD model for validation. While the basic MHD COCONUT simulations generally run on meshes with 1M to 1.5M cells, a grid of only 300k cells was used for these simulations since the AUSM+up scheme that is used for the MFMHD cases is more accurate than the HLL scheme for the basic ideal MHD set-up. This is well visible from  Figures~\ref{fig:Minimum_mhdvrrho_mfmhdvrrho} and \ref{fig:Maximum_mhdvrrho_mfmhdvrrho}, where the MHD (with HLL) and MFMHD (with AUSM+up) results are compared. In both these figures, the top row shows the MHD equivalent of the MFMHD simulation that is shown at the bottom, in other words on a mesh with 300k cells and with the same magnetogram and boundary conditions. In all cases, the bottom row, showing the MFMHD solution, shows many more detailed features in the solution compared to the top row MHD equivalent, both in the radial velocity plots (left) and the ion density plots (right). Higher detail is expected since the AUSM+up scheme is  less dissipative compared to HLL. 

To determine whether these features have a physical meaning, in the middle row a high-resolution low-dissipation MHD HLL solution is inserted, which was computed using enhanced magnetograms {(with five times higher magnetic field values) on a much finer grid (1.5M cells) and with less limiting, specifically designed to pronounce the electromagnetic features. While artificially enhancing the magnetic field strength might affect the exact topology of the resulting features, it still helps showcase well where they should roughly be located.} In both the minimum and maximum cases it is shown that, qualitatively, the AUSM+up features are indeed at the correct locations judging from these enhanced HLL simulations. 

{Apart from the resolution of the features due to the scheme, the different physics and formulation of the equations cause some variations in the large-scale features and velocity ranges.} When it comes to comparing the two MHD HLL solutions (the top and middle rows), a higher velocity is present in the refined resolution run (the middle row) as a result of stronger electromagnetic forces. In contrast, the MFMHD solutions (the bottom row) have a speed range  more similar to the equivalent ideal MHD set-up (the top row), with the values being lower due to the higher resistivity in the domain, as explained earlier. Overall, this shows the correct functioning of the developed ion-neutral COCONUT-MF solver, also when realistic data-driven applications are considered.

\subsection{Significance of charge exchange and collisional terms}

When comparing the MHD and MFMHD results above, it becomes clear that the MHD HLL solutions result in higher velocities and less developed structures. While the former is intuitive physically due to a higher resistivity in the domain thanks to the two-fluid Ohm's law formulation, the latter is likely also due to the applied numerical scheme, not just to the extended physics. For that reason, it is not possible to simply compare the MHD and MFMHD simulations to determine the effect of the coupling terms on the solution as the impacts of the scheme cannot be isolated. In order to evaluate the effects of the coupling terms, the simulations must be carried out with the same solver. 

Therefore, in total, thee MFMHD simulations were run for each case:

\begin{enumerate}
    \item the default set-up: collisions, ionisation/ recombination, and charge exchange included,
    \item collisional-only set-up: ionisation/ recombination, and charge exchange excluded, and
    \item coupling-free set-up: collisions, ionisation/ recombination, and charge exchange excluded.
\end{enumerate}

   \begin{figure*}
   \centering
   \includegraphics[width=15cm]{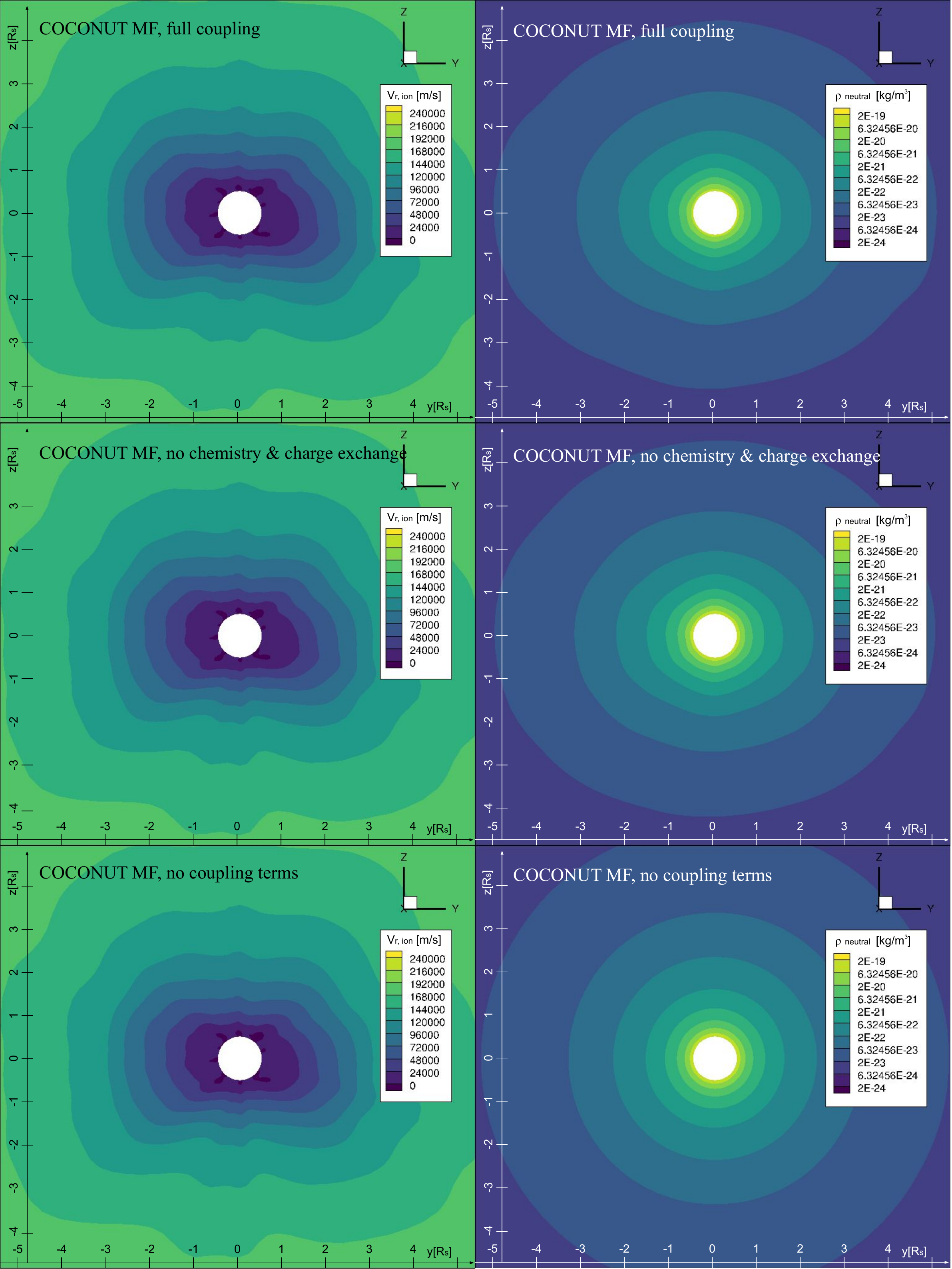}
   \caption{Three simulations of the solar minimum case are shown to help determine the effects of the different kinds of coupling terms. The top row shows the results for a full simulation (all coupling terms included). The middle row shows a run without charge exchange and chemical terms (but including collisional terms). Finally, the bottom row shows a run with no coupling terms. The left and right panels show the ion radial velocity and neutral density profiles, respectively. }
              \label{fig:Minima_Vrion_rhoneutral}%
    \end{figure*}

   \begin{figure*}
   \centering
   \includegraphics[width=15cm]{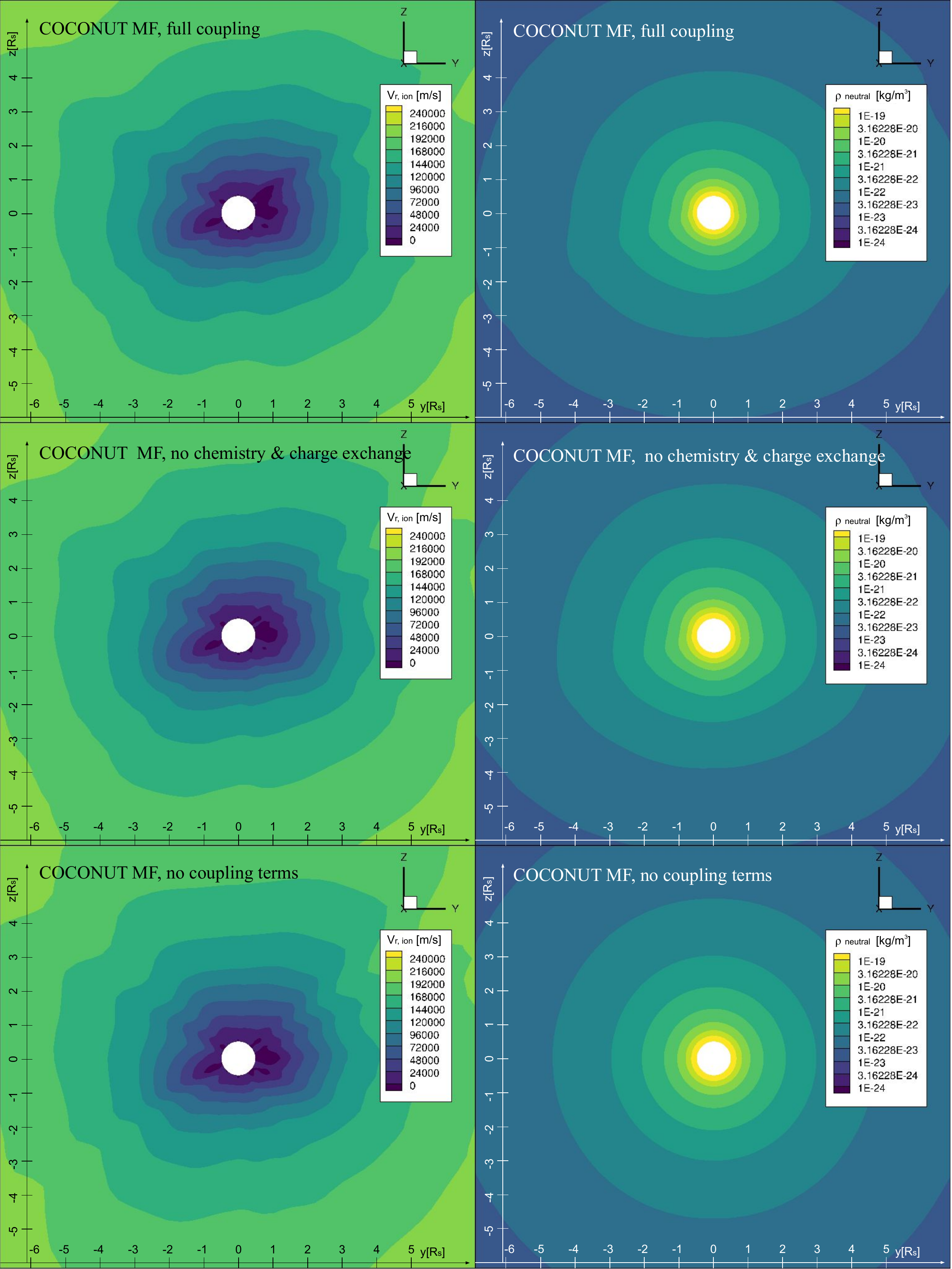}
   \caption{Three simulations of the solar minimum case are shown to help determine the effects of the different kinds of coupling terms. The top row shows the results for a full simulation (all coupling terms included). The middle row shows a run without charge exchange and chemical terms (but including collisional terms). Finally, the bottom row shows a run with no coupling terms. The left and right panels show the ion radial velocity and neutral density profiles, respectively. }
              \label{fig:Maxima_Vrion_rhoneutral}%
    \end{figure*}

The first was the default set-up, the results of which have already been presented above. The second set-up allows us to isolate the effects of the charge exchange and chemical terms. Finally, the third set-up helps us to understand the importance of the collisional terms separately and all the coupling terms together.

It should be noted that these terms were only handled in the fluid equations. The electric current remained to be computed from the full Ohm's law, since removing resistivity in the domain in the resistive MFMHD formulation would lead to divergence of the solver. Thus, it was not expected to retrieve the higher velocities observed in the ideal MHD results even with the coupling terms excluded. 

The results of these runs are shown in Fig.~\ref{fig:Minima_Vrion_rhoneutral} for the solar minimum case and Fig.~\ref{fig:Maxima_Vrion_rhoneutral} for the solar maximum case. The ion radial velocity is plotted (on the left) since the radial velocity field is where the different behaviour of the ions, due to changes in their dynamics, is generally the easiest to observe. Next, the neutral density (right) is also plotted since this state shows very well to what extent the fluids are still coupled and how the neutral fluid distributes itself in the domain as a result. 

These figures show that the ion radial velocities barely change when we exclude the coupling terms. The velocity field stays in the same range with only slight differences in the shape of the streamers in the maximum solar activity case. The plots also show that, if the charge exchange and chemical (ionisation and recombination) terms are removed, the neutrals change their concentration significantly but still follow the electromagnetic ion profiles. When the collisional terms are removed as well, the neutrals become decoupled and assume profiles that are spherically symmetric and determined by gravitational forces only.

The effects of the charge exchange and chemical terms also project onto the temperature profiles. For easier determination of these effects, the relative differences in the temperature of ions were computed for these three solutions as:

\begin{equation}
    dT_{\text{chem}} = \frac{T_{i, 2} - T_{i, 1}}{T_{i, 1}} \quad \hbox{and}\quad dT_{\text{col+chem}} = \frac{T_{i, 3} - T_{i, 1}}{T_{i, 1}},
\end{equation}

where the simulations with subscript $1$ refer to the default set-up (including all coupling terms), those with subscript $2$ refer to the set-up excluding only the chemical and charge exchange terms, and those with subscript $3$ refer to the runs without including any coupling terms in the fluid equations. 

\begin{figure*}
   \centering
   \includegraphics[width=15cm]{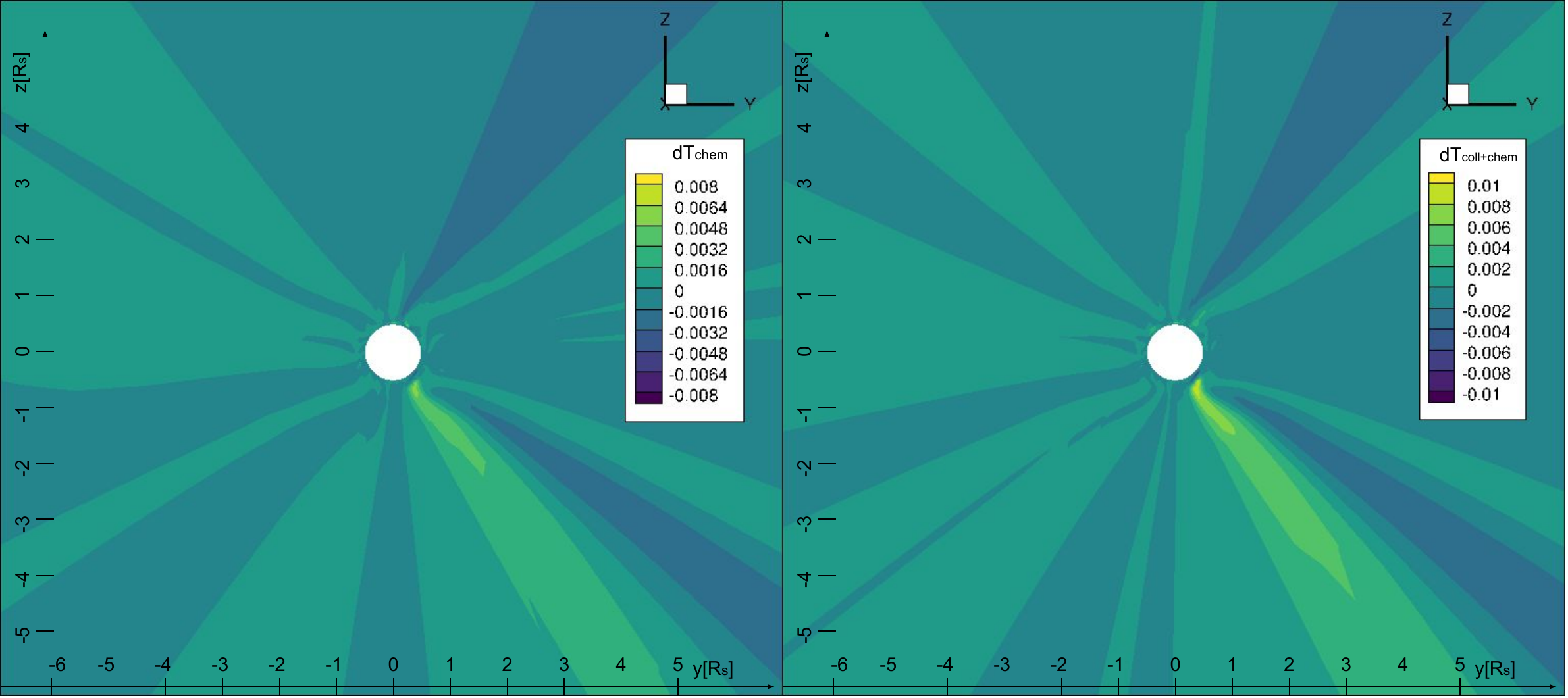}
   \caption{Relative differences in the temperature field after removing the charge exchange and chemical terms ($dT_{\text{chem}}$, left) and all coupling terms ($dT_{\text{col+chem}}$, right) when compared to the full default set-up for the case of the solar minimum.}
              \label{fig:Minimum_dT}%
    \end{figure*}

\begin{figure*}
   \centering
   \includegraphics[width=15cm]{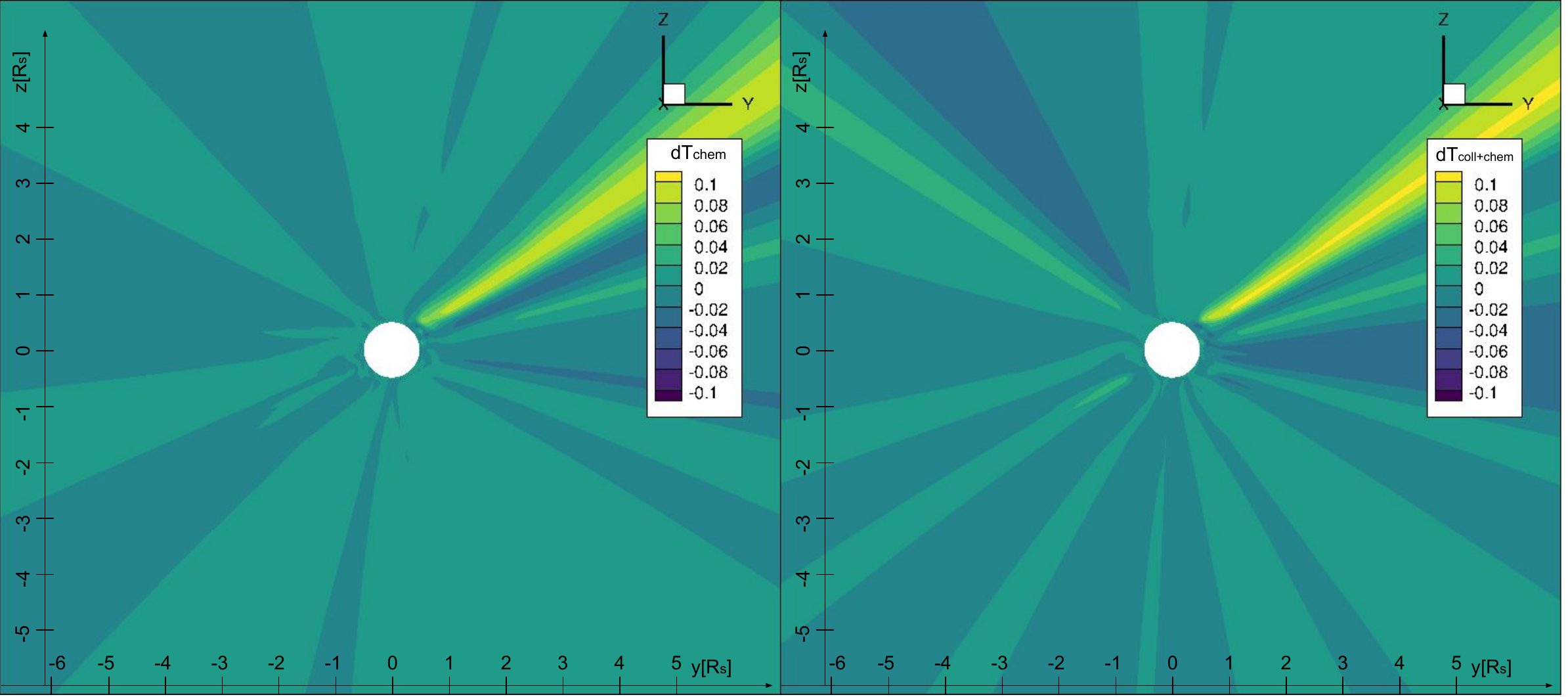}
   \caption{Relative differences in the temperature field after removing the charge exchange and chemical terms ($dT_{\text{chem}}$, left) and all coupling terms ($dT_{\text{col+chem}}$, right) when compared to the full default set-up for the case of the solar maximum. }
              \label{fig:Maximum_dT}%
    \end{figure*}

The results for $dT_{\text{chem}}$ and  $dT_{\text{col+chem}}$ for the solar minimum and maximum cases are shown in Figures~\ref{fig:Minimum_dT} and \ref{fig:Maximum_dT}, respectively. It is clear that, when it comes to the state of temperature, the charge exchange and chemical coupling terms do play a role. However, even then, these terms account for only up to 5\% and 10\% differences in the cases of the minimum and maximum of solar activity, respectively, and are generally enhanced only along the directions of the coronal streamers. {This is relevant, however, since these features especially are important for the correct resolution of the background solar wind for space weather forecasting. In other areas, the differences are minimal.}

From the perspective of correctly resolving the neutral fluid, naturally, all of the coupling terms are necessary since without them, as shown above, the concentration and distribution of the neutrals are not resolved accurately in the flow field. 

From the perspective of correctly resolving the ion fluid, the following observations can be made. Firstly, the exclusion of charge exchange, chemical, and collisional terms from the fluid equations does lead to effects on the ion radial velocity and temperature. Moreover, these effects are more pronounced (5 to 10\%) in the case of the solar activity maximum along the directions of coronal streamers. The reason why the effects are larger in the case of the solar maximum is most likely due to the fact that the presence of stronger electromagnetic features in the domain leads to higher gradients in velocity, density, and temperature, which affect the magnitude of the contributions of the respective terms. While these effects are of the order of a few percent, they are still much larger than what was expected considering the extremely small (1 to 1M) concentration of neutrals in the domain (given the usual assumption that the coronal is very close to being fully ionised). 

Secondly, the collisional terms are the most dominant coupling terms between the ion and the neutral fluids and, without these, the neutrals no longer respond to electromagnetic forces (and assume spherically symmetric profiles as shown in the bottom row). This shows that while the corona is frequently assumed to be collisionless there is enough collisionality to couple the ion and neutral fluids. 

Thirdly, the lower velocity reached in the domain of the MFMHD simulations when compared to MHD is due to the resistivity in Ohm's law, not due to the coupling terms in the fluid equations. This can be concluded judging from the fact that all of the simulations had the same velocity range, regardless of whether the coupling terms were included or not. 

While the results are surprising considering the low plasma density and the extremely low concentration of the neutrals, the offsets described above are still well within the ranges of uncertainties and offsets that are observed for ideal MHD simulations caused by, for example, the type or pre-processing of the magnetograms or the exact formulation of the inner boundary conditions. Thus, it is justified that the default COCONUT set-up runs with ideal MHD. 


These results do not concern, for example, ion-neutral waves or other time-dependent phenomena, since the solver determines only the steady-state solution. The results shown and discussed above also do not imply that the resolvable ion-neutral effects will remain negligible as the code is further advanced. For example, the addition of coronal mass ejections, creating strong shocks, might enhance the observed effects since we already saw that the increased complexity in the case of the solar maximum resulted in larger impacts compared to the minimum. Furthermore, once more physical terms are added (e.g. heating, thermal conduction, and radiation) and the domain is extended to capture the lower layers of the solar atmosphere, especially the chromosphere, the presence of the neutrals will be more significant and is expected to contribute considerably to the overall behaviour of the plasma. When that is implemented into our model, the utilisation of this two-fluid model will be of key importance.


%

\label{sec:performance}
   \begin{figure*}
   \centering
   \includegraphics[width=15cm]{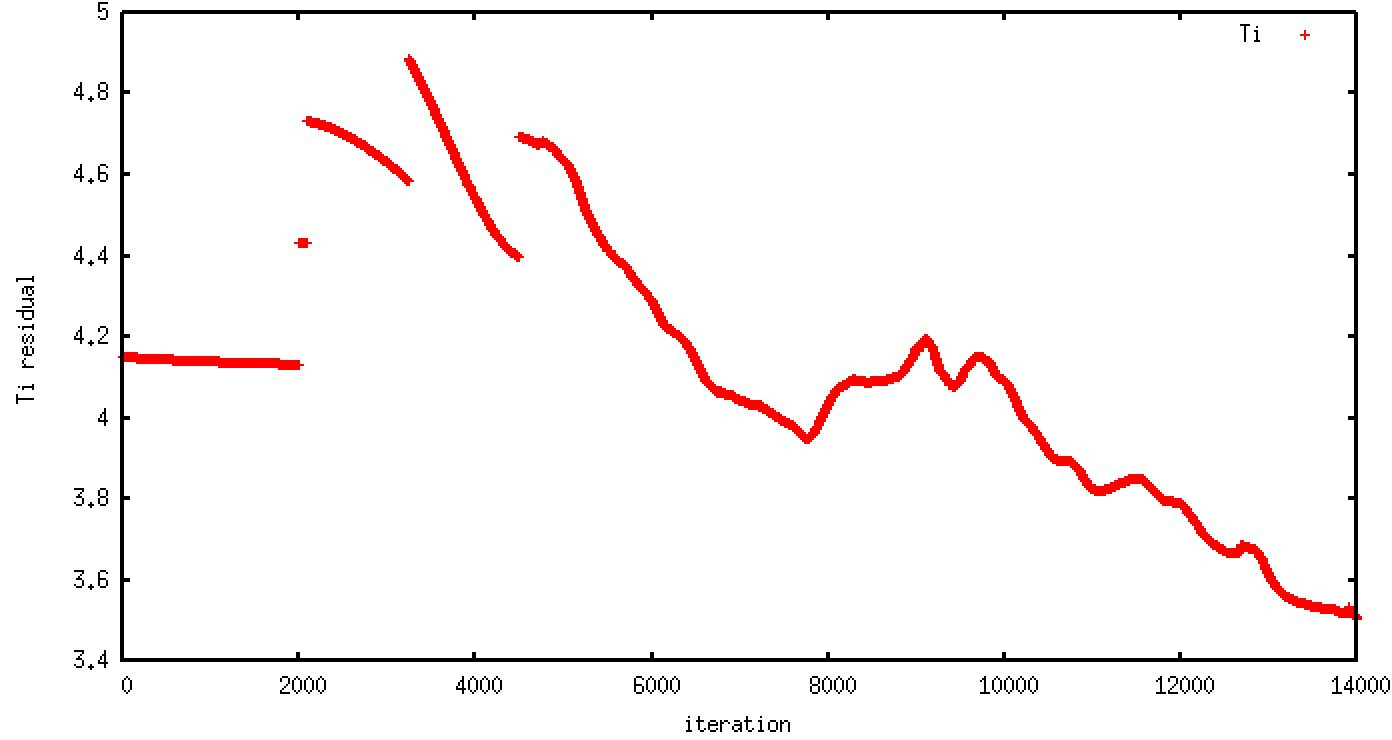}
   \caption{Evolution in the residual of ion temperature (as a logarithm) over the course of the run of the solar maximum case simulation.}
              \label{fig:Residual}%
    \end{figure*}

\section{Model performance}

While the MF solver allows us to capture more complex physics in the simulations (regardless of whether this physics influences the results significantly or not), the inclusion of an additional fluid in the set-up deteriorates the speed of the solver significantly. Each iteration takes longer compared to the ideal MHD solver due to the fact that instead of only nine equations being solved, now 18 are. In addition, due to the more complex physics involved, the problem is stiffer and the residuals take much longer to decrease to acceptable values compared to the ideal MHD solver. Additional stiffness of the solver also comes from the fact that it is solved dimensionally (as it was developed in the original reference \cite{ALVAREZLAGUNA2016252} where the solver is validated).

In Figure~\ref{fig:Residual}, we show an example residual evolution for a solar maximum simulation on a 300k cell grid, restarted from a converged MHD solution as described above. In this figure, since the equations are dimensional, the residual shows the logarithm of the changes between the subsequent steps in ion temperature in Kelvin. The jumps in the residual are due to an increasing CFL number, starting at two and then doubling with each increase (up to 32). In the end, the residual of 3.5 indicates roughly 3000$\;$K changes between the subsequent iterations:

\begin{equation}
    \mathrm{res}(T) = \log\sqrt{\sum_i\left(T_i^t - T_i^{t+1}\right)^2},
\end{equation}
where $i$ and $t$ are spatial and temporal indices. {It should be noted that the 3000$\;$K residual is summed up over the entire grid. Thus, the previous results showing local temperature differences of up to a few thousand Kelvin are physical since the actual cell residual is 3000$\;$K divided by the number of cells (300k), thus only around 0.01$\;$K. }

As can be seen in \cite{Brchnelova2022b}, on a mesh with 300k cells, the ideal MHD COCONUT solver only requires 2000 to 3000 iterations, while here, even after 14000 iterations, the residual is still not as low as the residual generally prescribed as a stop condition for the ideal MHD simulations (-3 for velocity as an adimensional value, which translates to roughly 2.7 in a dimensional setting). While the performance is still relatively good considering that the solution can be reached in a day using supercomputers (e.g. this specific case took around 1.5k CPU-hours), it would not make it possible to run such simulations operationally as a part of the space weather modelling toolchain, in which faster solutions at higher resolutions are needed. 

Potential numerical advancements planned for COCONUT in the future, such as adaptive mesh refinement (r-adaptivity, i.e. moving grids) combined with higher-order flux reconstruction, could solve this issue. Solving the MF equation in non-dimensional form could also help speed up its convergence because it would improve the conditioning number of the corresponding linearised system's matrix that is solved by the implicit code.  

\section{Conclusions}
\label{sec:conclusion}

In this work, we have introduced a MF extension of the COCONUT model called COCONUT-MF. This extension considers the ion and neutral species as separate fluids and takes into account their mutual interaction via coupling terms including collisions, ionisation, recombination, and charge exchange. Compared to the single-fluid ideal MHD formulation, it is also based on a less dissipative scheme, providing a higher accuracy on the same grid. The two aims of this paper were to validate the model/ solver and to use COCONUT-MF to determine the effects of the ion-neutral coupling terms on the global coronal modelling results. 

The COCONUT-MF model was first used to simulate a magnetic dipole for validation. The results were compared to the equivalent results obtained with the default ideal MHD COCONUT set-up. Both solutions showed a clear dipolar structure, with the specific shape depending on the scheme used (HLL in the case of COCONUT, AUSM+up in the case of COCONUT-MF). The COCONUT-MF results showed decreased velocities and increased temperature, which resulted from the fact that the MF formulation contains a higher resistivity in the domain.

Next, the new MF formulation was applied to two data-driven cases, namely a solar minimum case (using a magnetogram from August 1, 2008) and a solar maximum case (using a magnetogram from March 9, 2016). In both cases, it became clear that the AUSM+up scheme applied in the MF simulations led to more detailed structures on the same grid, making the velocity and density features more pronounced compared to the ideal MHD HLL simulations, as expected. Besides these differences, the simulations showed a qualitative agreement when it came to the distribution of streamers and density profiles just like in the case of the magnetic dipole. Similarly to the dipolar case, the resolved velocities were lower due to the increased resistivity. 

The significance of the coupling terms was evaluated next in order to quantify the effects of including the charge exchange, chemical, and collisional terms in the fluid equations on the plasma dynamics. This was done by running the MFMHD set-up three times for each case, once with the full physics included, once without the charge exchange and chemical terms, and once without all the coupling terms (i.e. also without the collisional contributions). The comparison between the profiles showed that despite their extremely small concentration the neutrals can still affect the dynamics of the plasma to a limited extent, namely up to 5 to 10\% (in terms of temperature) in the case of the solar maximum run in the regions along the coronal streamers. It was also demonstrated that sufficient collisionality exists in the domain to couple the ion and neutral fluids despite the low plasma density.

Finally, when evaluating the performance of the code, it was shown that the inclusion of the extra physics (by adding the charge exchange, chemical, and collisional terms to the equations) increased the stiffness of the problem and the complexity of the solver significantly, making the iterations more computationally expensive and the residual evolution less straightforward. Even with a relatively coarse grid, about 1.5k CPU-hours were required to obtain a solution. Considering that this added physics did not result in large changes in the solution for the given set-up, using this MF formulation for operational runs instead of the single-fluid MHD model in the current setting cannot be justified. 

Much work still must be done with COCONUT-MF in order to improve its performance, both in terms of physical accuracy and concerning convergence behaviour. The first step described in this paper was a basic formulation similar to the ideal MHD, polytropic global coronal model to give us insight into the effects of including a neutral species in a purely coronal set-up and resolving the simulation with a higher accuracy scheme. The next step would involve including even more physics that would be representative of the coronal heating mechanism(s), thermal conductivity, and radiative losses and an eventual extension to include the lower layers of the solar atmosphere, such as the chromosphere and the transition region, where the effect of the neutrals is expected to be very dominant because the ionisation degree is orders of magnitude lower there. To improve on the convergence rate, more advanced numerical techniques such as r-adaptation, in other words keeping the number of grid cells constant but moving them where needed (and thus coarsening the grid elsewhere), will eventually be implemented and a flux rope will be inserted to verify whether this code can eventually also resolve the dynamics of coronal mass ejections in the lower solar atmosphere just like in case of COCONUT \citep{Luis2023}.

\begin{acknowledgements}
      This work has been granted by the AFOSR basic research initiative project FA9550-18-1-0093. This project has also received funding from the European Union’s Horizon 2020 research and innovation programme under grant agreement No 870405 (EUHFORIA 2.0). These results were also obtained in the framework of the projects C14/19/089  (C1 project Internal Funds KU Leuven), G.0B58.23N  (FWO-Vlaanderen), SIDC Data Exploitation (ESA Prodex-12), and Belspo project B2/191/P1/SWiM. The resources and services used in this work were provided by the VSC (Flemish Supercomputer Centre), funded by the Research Foundation - Flanders (FWO) and the Flemish Government.
\end{acknowledgements}

%
%

\bibliographystyle{aa} 
\bibliography{biblio.bib}

\end{document}